\documentclass[showpacs,aps,prd,preprint,groupedaddress,preprintnumbers]{revtex4}

\usepackage{graphicx}
\usepackage{psfig}
\usepackage{hhline}

\def\2te{2{\theta}}

\def\'#1{\ifx#1i\accent19\i\else\accent19#1\fi}

\def\8{\infty}

\def\eps{\epsilon}

\begin{document}

\preprint{IFIC/05-65}

\title{Constraining non-standard interactions 
in $\nu_e e$ or $\bar{\nu}_e e$ scattering}

\author{J. Barranco}
\email{jbarranc@fis.cinvestav.mx}

\author{O. G. Miranda}
\email{Omar.Miranda@fis.cinvestav.mx}
\affiliation{Departamento de F\'{\i}sica, Centro de Investigaci{\'o}n y de 
  Estudios Avanzados del IPN, Apdo. Postal 14-740 07000 
  M\'exico, D F, M\'exico}

\author{C. A. Moura} 

\author{J. W. F. Valle} 
\email{valle@ific.uv.es} 
\affiliation{
  AHEP Group, Instituto de F\'{\i}sica Corpuscular --
  C.S.I.C./Universitat de Val{\`e}ncia \\
  Edificio Institutos de Paterna, Apt 22085, E--46071 Valencia, Spain}

\begin{abstract} We present a new analysis of non-universal and flavor
  changing non-standard neutrino interactions (NSI) in $\nu_e e$ or
  $\bar{\nu}_e e$ scattering.  Our global analysis of these process
  includes all relevant experiments, such as the most recent MUNU
  measurement from reactor neutrinos, both in the context of the
  Standard Model as well as extensions where NSI are present.  We also
  compare our constraints on non-universal and flavor changing NSI
  with results from previous analyses.
  We stress the importance of combining neutrino and anti-neutrino
  data in the resulting constraints on electroweak paramaters, and the
  important role that future low energy solar neutrino experiments can
  play in improving existing sensitivities.
\end{abstract}


\maketitle

\section{Introduction}
Electron-neutrino and electron-anti-neutrino scattering off electrons
have played an important role in the searches for neutrino
oscillations.  First hinted by the data from solar and atmospheric
neutrinos, oscillations have subsequently been confirmed with reactor
and accelerator data~\cite{fukuda:1998mi,ahmad:2002jz,eguchi:2002dm}.
Altogether, these experiments now give clear evidence that neutrinos
are massive~\cite{Maltoni:2004ei} and, therefore, expected to be
endowed with non-standard interactions that may violate leptonic
flavour and/or break weak universality~\cite{schechter:1980gr}.
Future experiments, such as BOREXINO~\cite{Alimonti:2000xc}, aim to
use the same reaction for detecting lower energy solar neutrinos.
The Standard Model cross section for this process has been known since
the 70's~\cite{Bardin:1970wr,'tHooft:1971ht,Chen:1972yi}, when the
first measurements have been carried out~\cite{Reines:1976pv}.
Radiative corrections have been calculated more recently
in~\cite{Bahcall:1995mm} and there have been recent
experiments~\cite{Auerbach:2001wg,Amsler:2002tu}.  Currently there are
many proposals to perform new experiments either at relatively high
energies~\cite{Conrad:2004gw}, in order to test the NuTeV
anomaly~\cite{Zeller:2001hh}, as well as at low
energies~\cite{Giomataris:2003pd,Giomataris:2003bp,Kopeikin:2003bx,neganov:2001bn},
motivated by the search for a possible non-zero transition neutrino
magnetic moment~\cite{Schechter:1981hw}.

As already mentioned, it has been long noticed that massive neutrinos
are expected to have non-standard interactions that may arise either
from the structure of the charged and neutral current weak
interactions in seesaw--type models~\cite{schechter:1980gr}.
Alternatively, they could arise from the exchange of scalar bosons, as
present in radiative and/or supersymmetric models of neutrino
mass~\cite{zee:1980ai,babu:1988ki}. The strength of the expected NSI
depends strongly on the model. Here we adopt a model independent
approach of simply analyzing their phenomenological implications in
neutrino electron scattering. For previous recent studies see
Refs~\cite{Berezhiani:2001rs,Berezhiani:2001rt,Davidson:2003ha}.

This possibility has been revived recently as it was noted that both
solar and atmospheric neutrino data are consistent with sizable values
of the NSI
parameters~\cite{Friedland:2004pp,Guzzo:2004ue,Friedland:2004ah}.  For
the case of neutrino interactions with the down--quark, it has been
shown that the presence of NSI brings in an ambiguous determination of
the solar neutrino oscillation parameters, with a new solution in the
so--called ``dark side'' (with $\sin^2\theta_{sol}\simeq
0.7$~\cite{Miranda:2004nb}), degenerate with the conventional one,
even after taking into account data from the KamLAND experiment. For
the case of $\nu_e e^-$ NSI the couplings are also allowed to be
large~\cite{Guzzo:2004ue}.

In this work we concentrate in the detailed study of $\nu_e e^-$ and
$\bar{\nu}_e e$ scattering in the presence of non-standard neutrino
interactions, which can not be found in previous studies, e.~g.
Refs.~\cite{Berezhiani:2001rs,Davidson:2003ha}.

We focus on short baseline terrestrial experiments such as the LSND
$\nu_e e^-$ scattering and a variety of $\bar{\nu}_e e$ scattering
experiments using reactor neutrinos, exploiting their complementarity.
Our analysis is new in two ways.  First we relax the conditions under
which the constraints on weak couplings have been previously derived.
Second, we update the study through the inclusion of more recent data,
such as the recent data from the MUNU
experiment~\cite{Daraktchieva:2003dr}.  Also for completeness, we
include the results from the Rovno reactor~\cite{Derbin:1993wy}.
Moreover, the results from the Irvine~\cite{Reines:1976pv} experiment
will be analyzed considering the two energy bins that were reported in
the original article.

This paper is organized as follows: in Sec.~\ref{sec:SM} we recall the
basics of $\nu_e e$ scattering in the context of the Standard Model,
in Sec.~\ref{sec:NSI} we analyse the role of non-standard neutrino
interactions and in Sec.~\ref{sec:res-fut} we discuss prospects for
further improvements, stressing the role of future low energy
experiments using solar neutrino, as well as experiments using
radioactive neutrino sources.
 
\section{The  neutrino electron scattering}
\label{sec:SM}

As a warm-up exercise, before considering the case of non-standard
neutrino interactions, let us briefly consider the restrictions placed
by current experiments within the context of the Standard Model.

\subsection{Preliminaries}
\label{sec:preliminaries}

In the Standard Model the $\nu_e e$ differential cross section
scattering involves both neutral and charged currents and is well
known~\cite{Bardin:1970wr} to be
\begin{equation}
\frac{d\sigma}{dT} = \frac{2 G_F m_e}{\pi}
\big[ 
g^2_L + g^2_R(1 - \frac{T}{E_\nu})^2 - g_L g_R \frac{m_e T}{E^2_\nu}
\big]
\label{diff:cross:sec}
\end{equation}
where $G_F=1.666\times10^{-5}$~GeV~$^2$, $m_e$ is the electron mass,
$T$ is the kinetic energy of the recoil electron and $E_\nu$ is the
neutrino energy. 

One can see explicitly that the differential cross section in
Eq.~(\ref{diff:cross:sec}) has a symmetry under the simultaneous
transformation $g_L\to -g_L$ and $g_R\to -g_R$. Apart from the last
term, it is also invariant under separate sign changes in $g_{L,R}$.
For the case of $\bar{\nu}_e e$ scattering one has to exchange $g_L$
by $g_R$.

For a fixed neutrino energy, the determination of the weak coupling
constants $g_L-g_R$, is ambiguous since the same cross section in
Eq.~(\ref{diff:cross:sec}) is achieved for any $g_L-g_R$ values in an
ellipse with one axis given by $1$ and the other one by $(1 -
\frac{T}{E_\nu})$.
However, measurements at different neutrino energies can potentially
lift this degeneracy, due to the last term in
Eq.~(\ref{diff:cross:sec}).  For example, for sufficiently low
energies, comparable to the electron mass, the extra term rotates the
ellipse by a sizable angle
\begin{equation}
\tan 2\theta = 
\frac{m_e}{(2 E_\nu - T)}.
\end{equation}

On the other hand, the anti-neutrino cross section defines another
ellipse which is perpendicular to the one corresponding to the neutrino
case, since the axis width of this ellipse is exactly opposite ($g_L
\leftrightarrow g_R$).  Therefore, by judicious combinations of
energies and/or adding anti-neutrino data, one expects to lift the
above degeneracy, as we will see in the next subsection.

Within the Standard Model the coupling constants $g_L$ and $g_R$ are
expressed, at tree level, as
\begin{eqnarray}
\label{gLgR}
g_L &=& \frac12 + \sin^2\theta_W\\
g_R &=&    \sin^2\theta_W
\end{eqnarray}
where $g_L \equiv 1+g_L^{SM}$, $g_L^{SM}$ being the conventional SM
definition.  We have checked explicitly that for the present accuracy
of the experiments, the above simple formulae are sufficient, as there
is no sensitivity to the corresponding radiative corrections given
in~\cite{Bahcall:1995mm}.

\subsection{Analysis}
\label{sec:analysis}

In our global analysis of the $\nu_e e$ and $\bar{\nu}_e e$ scattering
we have included all current experiments, namely, the data from the
LSND measurement of the neutrino electron scattering cross
section~\cite{Auerbach:2001wg}; for the anti-neutrino electron
scattering we have considered the two bins measured in the Irvine
experiment~\cite{Reines:1976pv}, the results of the Rovno
experiment~\cite{Derbin:1993wy} and the more recent result from the
MUNU experiment~\cite{Daraktchieva:2003dr}. The experimental results
are summarized in Table~\ref{table:1}.

In order to perform the analysis we need the total cross section which, 
for the antineutrino case  we express as
\begin{equation}
\sigma =  \int dT' \int dT \int dE_\nu 
                    \frac{d\sigma}{dT} \lambda (E_\nu) R(T,T')
\label{diff:cross:sec-sm}
\end{equation}
where  both spectra and the detector energy resolution function,
should be convoluted with the cross sections given in
Eq.~(\ref{diff:cross:sec}).

In particular for the most recent MUNU measurement from reactor
neutrinos~\cite{Daraktchieva:2003dr}, we use an anti-neutrino 
energy spectrum given by 
\begin{equation}
\lambda(E_nu) = \sum_{k=1}^4 a_k \lambda_k(E_\nu) 
\label{diff:cross:sec-NSI}
\end{equation}
where $a_k$ is the abundance of $^{235}$~$U$ ($k=1$), $^{239}$~$Pu$
($k=2$), $^{241}$~$Pu$ ($k=3$) and $^{238}$~$U$ ($k=4$) in the
reactor, $\lambda_k(E_\nu)$ is the corresponding neutrino energy
spectrum which we take from the parametrization given
in~\cite{Huber:2004xh}, with the appropriate fuel composition. For
energies below $2$~MeV there are only theoretical calculations for the
antineutrino spectrum which we take from Ref.~\cite{Kopeikin:1997ve}.
For the case of the Irvine experiment we prefer to use the neutrino
energy spectrum used by the experimentalists at that
time~\cite{Avignone}.

Regarding the detector resolution function $R(T,T')$ for the case of
MUNU it was found to be 8 \% scaling with the power $0.7$ of the
energy~\cite{Daraktchieva:2003dr}.
For the other two anti-neutrino experiments included in our analysis
the resolution function was not reported, so we neglect resolution
effects. 

\begin{table}[!t] 
\begin{tabular}{|c|c|c|c|}
\hline
Experiment & Energy range (MeV) & events & measurement \\
\hline\hline 
LSND $\nu_e e$   & 10-50  & 191 &
$\sigma=[10.1\pm1.5]\times 
E_{\nu_e}({\rm MeV})\times10^{-45} {\rm cm}^2$   \\[.1cm]
Irvine $\bar{\nu}_e-e$  & 1.5 - 3.0  & 381 &
$\sigma=[0.86\pm0.25]\times \sigma_{V-A}$   \\[.1cm]
Irvine $\bar{\nu}_e-e$   & 3.0 - 4.5 & 77 &
$\sigma=[1.7\pm0.44]\times \sigma_{V-A}$   \\[.1cm]
Rovno $\bar{\nu}_e-e$  & 0.6 - 2.0 & 41 &
$\sigma=(1.26\pm 0.62)\times10^{-44} {\rm cm}^2 / {\rm fission}$ \\[.1cm]
 MUNU $\bar{\nu}_e-e$  & 0.7 - 2.0  & 68 &
$1.07 \pm 0.34$ events day~$^{-1}$  \\[.1cm]
\hline
\end{tabular}
\caption{Current experimental data on (anti-)neutrino electron scattering.}
\label{table:1}
\end{table}

For the LSND electron neutrino experiment we use the theoretical
expectation for the total neutrino electron cross section, which is 
\begin{equation}
\sigma(\nu_e e) = \frac{2m_e G_F^2 E_\nu}{\pi}[g_L^2 + \frac13 g_R^2]. 
\end{equation}
Notice that in this case the term $g_L g_R$ can be neglected, since
this experiment was done at high energies of tens of MeV. As a result
there is no tilt in the ellipse, as discussed in the previous section
(see also Fig.~\ref{fig:global}).

With this information we proceed to our $\chi^2$ analysis.
Altogether, we will have five observables, and therefore, it will be
possible to constrain up to four parameters simultaneously.
We neglect correlations between experiments; this is a good
approximation as the only possible correlation comes from the reactor
neutrino energy spectrum, estimated to be less than
2\%~\cite{Huber:2004xh}, small in view of the statistical errors.
Therefore we can define the $\chi^2$ simply as
\begin{equation}
\chi^2 = \sum_i\frac{(\sigma_i^{\rm theo}-\sigma_i^{\rm exp})^2}{\Delta_i^2}
\end{equation}
where the $\sigma_i^{\rm exp}$ are given by the measurements shown in
Table~\ref{table:1} and $\Delta_i$ are the corresponding errors, while
$\sigma_i^{\rm theo}$ is the theoretical expectation. 

\subsection{The Standard Model parameters}

In this section we present the results of our fit first in terms of
the $g_L$ and $g_R$ coupling constants and, later, we will obtain the
value of the Standard Model weak mixing angle.

To obtain the allowed regions for the $g_L$ and $g_R$ coupling
constants we perform a $\chi^2$ analysis as discussed in the previous
subsection. These two parameters are determined by five measurements
and therefore we will have three degrees of freedom.  The minimun
$\chi^2$ for this case was $0.52$.

\begin{figure}
\begin{center}
\includegraphics[width=0.6\textwidth,angle=-90]{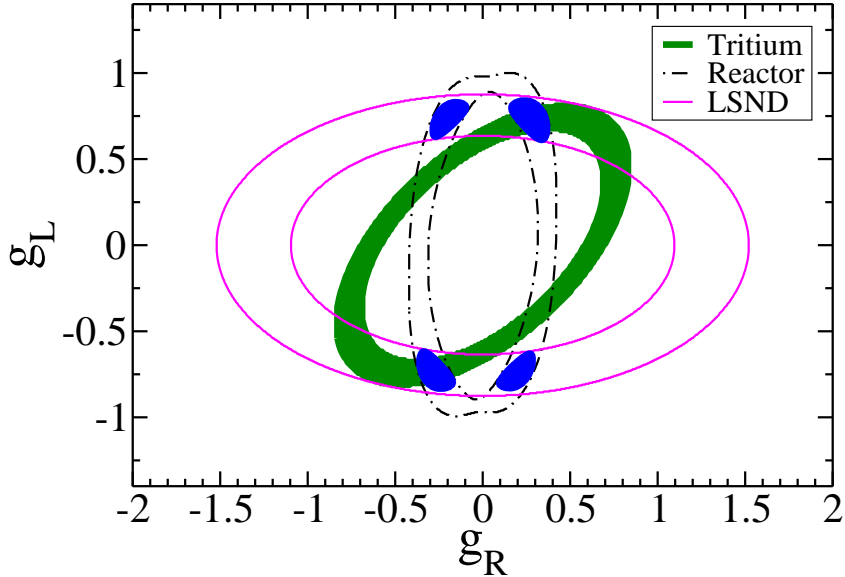}
       \caption{
         Allowed $90$ \% C. L. regions for $g_L$ and $g_R$ obtained by
         a global fit to neutrino and anti-neutrino electron
         scattering data. It is possible to see the existence of four
         allowed regions. The plot also shows the contribution from
         LSND neutrino electron scattering (horizontal ellipse) and
         combined data from reactor experiments (vertical
         ellipse). The tileted ellipse illustrates the potential of a
         future low-energy artificial neutrino source (Tritium
         proposal in Ref.~\cite{Giomataris:2003bp}).  }
        \label{fig:global}
\end{center}
\end{figure}

The results are illustrated in Fig.~(\ref{fig:global}) for $90$ \% C L
($\Delta \chi^2 =$ $4.61$). In this case one can clearly notice
the existence of four possible regions for these parameters.
We overlay in the same figure the corresponding equi-cross section
regions for current neutrino and anti-neutrino experiments, which form
two perpendicular ellipses, as expected.
The neutrino LSND data gives rise to the horizontal ellipse, while the
combined anti-neutrino data lead to the vertical ellipse and therefore
reduce the allowed region by restricting the $g_L$ and $g_R$ values to
the intersection of the two.
Of the existing experiments the ones giving the main contribution to
the constraint are the LSND and the Irvine experiments, due to their 
higher statistics. A more restrictive analysis from the MUNU
experiment might be possible by using its binned data, although this
is out of the scope of the present work.

We also show in Fig.~(\ref{fig:global}) the case of a future
low-energy neutrino experiment, in which case the ellipse is tilted.
To illustrate the potential of future low energy experiments we
consider, for definiteness, the case of the NOSTOS proposal, where
antineutrinos come from an intense Tritium source with a maximum
energy of 18.6 KeV~\cite{Giomataris:2003pd}.
For this case the anti-neutrino spectrum for the source is taken
as~\cite{Ianni:1999nk}
\begin{equation}
\lambda \left(E_{\nu}\right)=
A\frac{x}{1-e^{-x}}\left(Q+m_{e}-E_{\nu}\right) E_{\nu}^{2}
\sqrt{\left(Q+m_{e}-E_{\nu}\right)^{2}-m_{e}^{2}}
\, ,\label{lambda}
\end{equation}
where $A$ is a normalization factor, $Q=18.6$~KeV, $m_{e}$ is the
electron mass, and
\begin{equation}
x=2\pi \alpha_{\rm e.m.} \frac{Q+m_{e}-E_{\nu}}
{\sqrt{\left(Q+m_{e}-E_{\nu}\right)^{2}-m_{e}^{2}}} \,
.\label{lambda2}
\end{equation}
This spectrum is convoluted with the anti-neutrino differential cross
section. The total number of events is set to be
$3500$~\cite{Giomataris:2003bp} (for one year of data taking).

We see that there is room for such future low-energy neutrino
experiments to provide useful input to resolve the current degenerate
determination of the weak coupling constants, improving the existing
measurements.
Unfortunately, as discussed above, the symmetry of the cross section
when we make the transformations $g_L\to -g_L$ and $g_R\to -g_R$ can
not be lifted by this method.  Such a degeneracy is therefore
irreducible. This is not an academic ambiguity as it means the
validity of the gauge theory description dictated by the Standard
Model. In order to test the future sensitivity we set the experimental
measure to be exactly the SM prediction, and we consider only the
statistical error. After considering this experimental set up for
NOSTOS we obtain the region shown in figure 1.

Assuming the validity of the Standard Model, given by
Eqs.~(\ref{gLgR}) our results can also be presented directly in terms
of the weak mixing angle. In this case, the combined analysis of the
existing (anti)-neutrino-electron scattering experiments gives
$\sin^2\theta_W=0.27\pm 0.03$.  The corresponding minimum for the
$\chi^2$ function was $\chi^2_{\rm min}= 0.89.$
\begin{figure}
\centering
\includegraphics[width=0.6\textwidth,angle=-90]{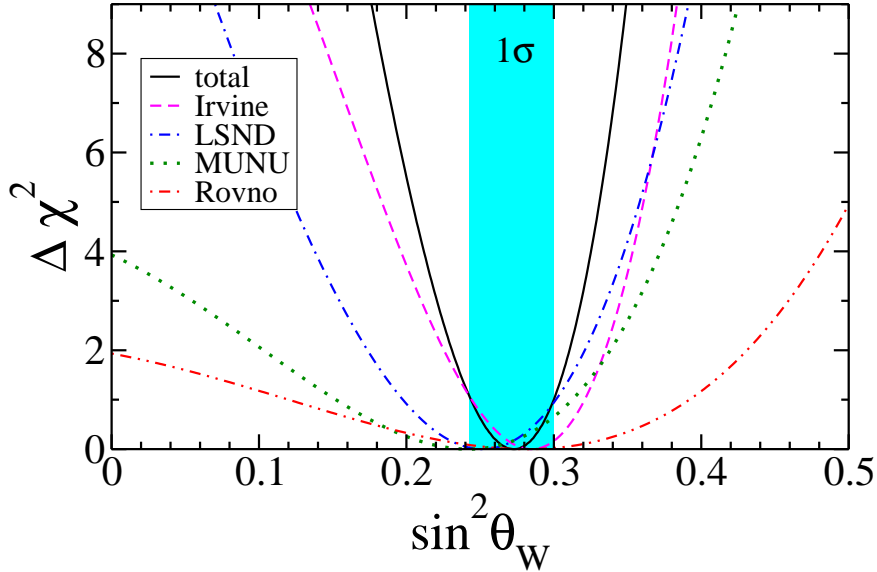}
       \caption{
         $\Delta \chi^2$ for $\sin^2 \theta_W$ from $\nu_e e$ or
         $\bar{\nu}_e e$ scattering. The contribution of each
         experiment to the $\Delta \chi^2$ is also shown. 
       }
        \label{fig:sin}
\end{figure}

The various contributions to $\Delta \chi^2$ from different individual
experiments are indicated in Fig.  \ref{fig:sin}. Note that the
present fit gives a central value higher than the world
average~\cite{Eidelman:2004wy}, though the error is larger than found
in other experiments, because of their small statistics relative to
collider experiments.  Nevertheless we find this to be interesting as
an independent and clean probe of the Standard Model.

\section{Non-standard interactions in $\nu_e e$ and $\bar{\nu}_e e$ scattering}
\label{sec:NSI}

Solar neutrino data are robust with respect to possible modifications
in solar physics involving various types of magnetic fields both in
the convective zone~\cite{Miranda:2004nz} as well as radiative
zone~\cite{Burgess:2003fj,Burgess:2003su}. If present, non-standard
effects are expected to be sub--leading insofar as providing an
explanation of the existing data~\cite{pakvasa:2003zv}.  However, even
taking into account the crucial data from reactor experiments, the
current accepted interpretation of solar neutrino data is not yet
robust when neutrinos are endowed with non-standard
interactions~\cite{Miranda:2004nb}. In fact it has been shown that the
presence of NSI brings in an ambiguous determination of the solar
neutrino oscillation parameters, with a new ``dark side'' solution
(with $\sin^2\theta_{sol}\simeq 0.7$~\cite{Miranda:2004nb}),
essentially degenerate with the conventional one.  Similarly, despite
the good description provided by oscillations of contained and upgoing
events which leads to limits on the strength of the NSI strength in a
two--neutrino scenario~\cite{fornengo:2001pm}, atmospheric neutrino
data are still consistent with sizable values of the NSI parameters
when three neutrinos are considered~\cite{Friedland:2004ah}.
Here we focus on the case of terrestrial experiments involving
electron--type neutrinos and anti-neutrinos.

\subsection{Cross section}
\label{sec:cross-section}

A model independent way of introducing such non standard interactions
is via the effective four fermion Lagrangian \cite{Berezhiani:2001rs}
\begin{equation}
-{\cal L}^{eff}_{\rm NSI} =
\varepsilon_{\alpha \beta}^{fP}{2\sqrt2 G_F} (\bar{\nu}_\alpha \gamma_\rho 
L
\nu_\beta)
( \bar {f} \gamma^\rho P f ) \label{lagrangian_nsi}
\end{equation}
where $f$ is a first generation SM fermion: $e,u$ or $d$, and $P=L$ or
$R$, are chiral projectors.  With this Lagrangian (\ref{lagrangian_nsi})
added to the Standard Model Lagrangian one can compute the
differential cross section for the process $\nu_e e\to \nu_\alpha e$ as

\begin{eqnarray}\label{cross-section}
\lefteqn{{d\sigma(E_{\nu}, T) \over dT}= {2 G_F^2 M_e \over \pi} [ (\tilde 
g_L^2+\sum_{\alpha \neq e}
|\epsilon_{\alpha e}^{e L}|^2)+{} } \nonumber\\ & & {}+
(\tilde g_R^2+\sum_{\alpha \neq e}
|\epsilon_{\alpha e}^{e R}|^2)\left(1-{T \over E_{\nu}}\right)^2-
(\tilde g_L \tilde g_R+ \sum_{\alpha \neq e}|\epsilon_{\alpha e}^{e L}||
\epsilon_{\alpha e}^{e R}|)m_e {T \over E^2_{\nu}}]
\end{eqnarray}
with $\tilde g_L=g_L+\epsilon_{e e}^{e L}$ and $\tilde
g_R=g_R+\epsilon_{e e}^{e R}$. This equation has six NSI parameters,
two of them correspond to non-universal (NU) NSI: $\epsilon_{ee}^{e
  LR}$ and four to flavor changing (FC) NSI: $\epsilon_{e\mu}^{e LR}$
and $\epsilon_{e\tau}^{e LR}$. In view of the stringent (though
indirect) constraints on the FC parameters $|\epsilon_{e\mu}^{e
  LR}|<7.7\times10^{-4}$~\cite{Davidson:2003ha} we will, for
simplicity, neglect FC NSI involvion muon neutrinos. This way we are
left with the two NU NSI parameters and two FC parameters,
$\epsilon_{e\tau}^{e LR}$.

The agreement between $\nu_e e$ scattering experiments and the
Standard Model predictions had been previously studied in
Ref.~\cite{Berezhiani:2001rs,Berezhiani:2001rt,Davidson:2003ha} in
order to place restrictions on the magnitude of non-standard
interactions.
However, existing analyses either restricted the variation of the
parameters, which were considered only
one--at--a--time~\cite{Davidson:2003ha}, or the combination of two NSI
parameters (the non-universal coupling $\epsilon_{e e}^{e R}$ and
$\epsilon_{e e}^{e R}$) but using only two
experiments~\cite{Berezhiani:2001rs}.

Here we revisit this question generalizing the conditions under which
these constraints have been derived and, as we have already mentioned,
updating the study through the inclusion of more recent data, such as
the data from the MUNU experiment~\cite{Daraktchieva:2003dr}. Also for
completeness, we will consider the results from the Rovno
reactor~\cite{Derbin:1993wy}.  Moreover, the results from the Irvine
experiment will be analyzed considering the two energy bins that were
reported in the original article.

Although the constraints are expected to be weaker in our case, they
will be robust than the ones obtained in the case where the parameters
are taken only one--at--a time in the analysis.
However, as will be clear at the end of this section, by taking full
advantage of the combination of neutrino and anti-neutrino data we are
able to obtain more stringent bounds on ``right-handed'' NSI
parameters than previously.

\subsection{NSI Analysis}
\label{sec:nsi-analysis}

\begin{figure}
\centering
\includegraphics[width=0.6\textwidth,angle=-90]{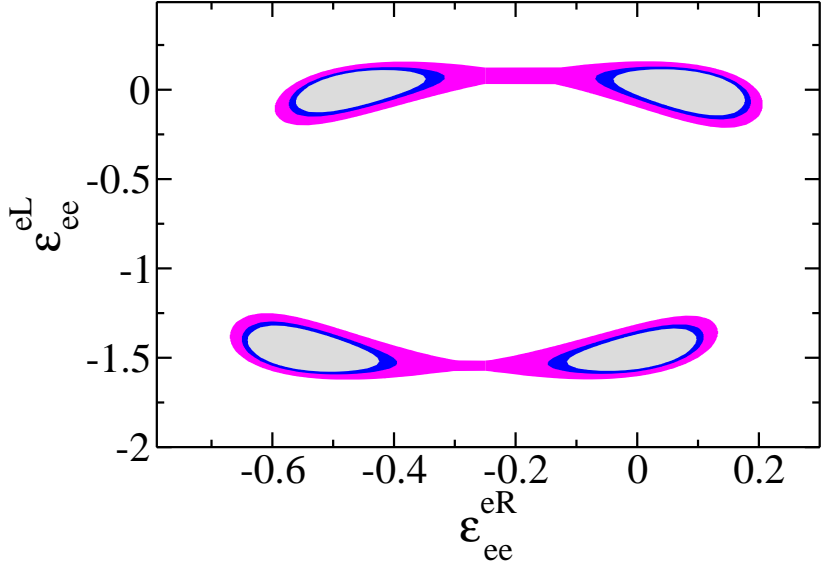}
       \caption{
Allowed regions at 90, 95 and 99 \% C. L. for
$\varepsilon^{eL}_{ee}$ and $\varepsilon^{eR}_{ee}$ obtained by a global fit
to neutrino and anti-neutrino electron scattering. The flavor changing 
NSI parameter were taken equal to zero.
}       
\label{fig:NU-NSI-2par}
\end{figure}

First we present the results for the case of non-universal NSI
($\epsilon^{eL}_{ee}$, $\epsilon^{eR}_{ee}$ ), with flavour changing
parameters set to zero. In Fig. (\ref{fig:NU-NSI-2par}) we show the
allowed regions at 90, 95 and 99 \% C L ($\Delta \chi^2 =$~$4.61$,
$5.99$, $9.21$).  The minimum $\chi^2$ was 0.52.
One can see that its determination is improved with respect to the
current results, although a twofold ambiguity in $\epsilon^{eL}_{ee}$
remains.
This follows from the discussion given in section \ref{sec:SM}, where
we stressed that the intersection from the neutrino and anti-neutrino
ellipses (see Fig. \ref{fig:global}) does not allow for a unique
discrimination of the coupling constant values. It is here that future
low energy experiments have a chance of improving their determination.

The same analysis can be performed for the case where we allow only
flavor changing NSI parameters ($\epsilon^{eL}_{e\tau}$,
$\epsilon^{eR}_{e\tau}$ ), or for the general case when we take into
account all four parameter simultaneously. The results of this
analysis are summarized in table \ref{table:results}. The left column
collects previously reported constraints~\cite{Davidson:2003ha},
determined under the assumption that only one NSI parameter was
allowed to take on nonzero values. In the second column, for
comparison, we present the result of our fit for the same case of a
one-parameter analysis. The third column gives our result for a two
parameters analysis, where only NU or FC parameters are non-zero,
therefore the NU region corresponds to the one shown in
Fig.~\ref{fig:NU-NSI-2par}.  Finally, the fourth column shows a more
general case when from the four parameters we take a projection over
two of them (either NU or FC) allowing the other two to take on
non-zero values. In this case for a 90 \% C L we have again to
consider $\Delta\chi^2=4.61$ but the regions are wider as can be seen
from the table.  The minimum $\chi^2$ for this analysis was 0.49.

One can see that the constraints for the case when only one parameter
is considered are similar to the results previously
reported~\cite{Davidson:2003ha}, with the exception of
$\eps_{e\tau}^{eR}$ and $\eps_{ee}^{eR}$ where ours are clearly
better.  This is natural to expect and follows from the fact that we
are combining the LSND neutrino electron scattering data with the
anti-neutrino electron scattering data. This allows us to obtain four
different regions for the left and right couplings as can also be seen
from Fig.~(\ref{fig:global}).
It is important to note, however that when the four parameters are
taken as freely-varying our constraints are weaker than the existing
ones for the case of ``left-handed'' couplings $\eps_{e\tau}^{eL}$ and
$\eps_{ee}^{eL}$, as expected (in fact they could be as large as order
unity).  In contrast, for the case of the ``right-handed'' NSI
couplings, our constraints better than the previous limits. The
explanation of this puzzle is that, in this case, in contrast to
previous work, we combine neutrino and anti-neutrino data. As we have
already seen, this has a great impact in constraining the
``right-handed'' NSI paramneters.

\begin{table}[!t] 
\begin{tabular}{|c|c|c|c|c|}
   \hline      {\rule[-3mm]{0mm}{8mm}  } & Previous Limits  & One parameter &
            Two Parameters & All Parameters 
            \\ \hline \hline
$\eps_{ee}^{eL}$ &$-0.07 < \eps_{ee}^{eL} < 0.11  $ 
                  & $-0.05 < \eps_{ee}^{eL} < 0.12 $ 
                    & $-0.13< \eps_{ee}^{eL} < 0.12$ 
                     & $ -1.58 < \eps_{ee}^{eL} < 0.12$   \\ [.1cm]
    \hhline{|-|-|-|||}
   $\eps_{ee}^{eR}$ & $-1.0 < \eps_{ee}^{eR} < 0.5 $ 
                     & $-0.04 < \eps_{ee}^{eR} < 0.14 $ 
                       & $-0.07< \eps_{ee}^{eR} < 0.15 $
                        & $-0.61 < \eps_{ee}^{eR} < 0.15$   \\ [.1cm]
     \hline
   $\eps_{e\tau}^{eL}$ & $ |\eps_{e\tau}^{eL}| < 0.4 $ 
                        & $ |\eps_{e\tau}^{eL}| < 0.44 $ 
                         &$ |\eps_{e\tau}^{eL}| < 0.43$ 
                           & $|\eps_{e\tau}^{eL}| < 0.85$   \\ [.1cm]
    \hhline{|-|-|-|||}
   $\eps_{e\tau}^{eR}$ &$ |\eps_{e\tau}^{eR}| < 0.7 $ 
                        & $ |\eps_{e\tau}^{eR}| < 0.27 $ 
                         & $ |\eps_{e\tau}^{eR}| < 0.31 $
                          & $ |\eps_{e\tau}^{eR}| < 0.38$   \\ [.1cm]
            \hline
        \end{tabular}  
    \caption{Constrains on NSI parameters at 90 \% C L. In the first
 column we show the previous constraints obtained
 in~\cite{Davidson:2003ha}, while in the second we show the
 corresponding results found in the present analysis. The last two
 columns show the case in which two and four parameters are allowed to
 vary simultaneously (see the text for explanation).}
    \label{table:results} 
\end{table}

\section{Summary and prospects}
\label{sec:res-fut}

We have presented a global analysis of non-standard neutrino
interactions in electron (anti)-neutrino scattering off electrons,
including all current experiments, such as the most recent MUNU
measurement from reactor neutrinos. We have discussed the resulting
constraints both in the context of the Standard Model as well as
extensions where non-standard neutrino interactions are present. We
obtained constraints on non-universal and flavor changing NSI and
compared our bounds with those obtained in previous analyses.
We find that substantial room for improvement is expected from
$\nu_e e$ or $\bar{\nu}_e e$ low-energy scattering experiments.
There are several proposals of this type, either using solar
neutrinos, such as BOREXINO~\cite{Alimonti:2000xc}, or experiments
using artificial neutrino sources, such as \cite{neganov:2001bn}, that
will be helpful in constrining NSI parameters as well as for other
type of new physics (see for
example~\cite{barabanov:1998bj,miranda:1998vs} ). From the point of
view of pinning down the interactions of the $\nu_e$ and $\bar{\nu}_e$
low energy scattering experiments offer an alternative frontier that
complements information that comes from higher
energies~\cite{Chooz,Gouvea}.

In summary, cross section measurements by themselves, at a given
energy, lead to a degeneracy in the coupling constants and, therefore,
in the determination of the NSI parameters. This degeneracy can be
partially removed by considering both neutrino and anti-neutrino
scattering off electrons. Further improvements require low energy
neutrino experiments.

\acknowledgments We thank Nicolao Fornengo for reading the manuscript.
This work has been supported by Spanish grant BFM2002-00345,
Conacyt-M\'exico, and by the EC RTN network MRTN-CT-2004-503369.
C. A. M.  is supported by AlBan Scholarship
no. E04D044701BR. J. B. would like to thank to IFIC/CSIC for the kind
hospitality during the visit where part of this work was done.

\end{document}